\documentclass[aps,prb,twocolumn,floatfix,superscriptaddress,goodfloat]{revtex4}
\usepackage{epsf}
\usepackage{isolatin1}
\usepackage{graphics}
\usepackage{bbm}
\newcommand{\R}{{\mathbbm{R}}}

\newcommand{\myscalebox}[1]{\scalebox{0.35}[0.35]{#1}}

\begin{document}
\title{Ground state of the Bethe-lattice spin glass and
  running time of an exact optimization algorithm }

\author{Frauke Liers} \email{liers@informatik.uni-koeln.de}
\affiliation{Universität zu Köln, Institut für Informatik,
  Pohligstraße 1, 50969 Köln, Germany}

\author{Matteo Palassini} \email{matteo@maxwell.ucsf.edu}
\affiliation{University of California San Francisco, 3333 California
  Street, Suite 415, San Francisco, CA 94118, USA}

\author{Alexander K. Hartmann} 
\email{hartmann@theorie.physik.uni-goettingen.de}
\affiliation{Institut für Theoretische Physik, Universität Göttingen,
  Bunsenstr. 9, 37073 Göttingen, Germany}

\author{Michael Juenger} \email{mjuenger@informatik.uni-koeln.de}
\affiliation{Universität zu Köln, Institut für Informatik,
  Pohligstraße 1, 50969 Köln, Germany} \date{\today}
\begin{abstract}
  We study the Ising spin glass on random graphs with fixed
  connectivity $z$ and with
  a Gaussian distribution of the couplings, with mean $\mu$ and unit
  variance.  We compute exact ground states by using a sophisticated
  branch-and-cut method for $z=4,6$ and system sizes up to 1280 spins,
  for different values of $\mu$.
  We locate the spin-glass/ferromagnet phase transition at
  $\mu =  0.77 \pm 0.02$ ($z=4$) and $\mu = 0.56 \pm 0.02$ $(z=6)$.
  We also compute the energy and magnetization in the Bethe-Peierls
  approximation with a stochastic method, and estimate the magnitude
  of replica symmetry breaking corrections.
Near the phase transition, we observe a sharp change of the
median running time  of our implementation of
the algorithm, consistent with a change
from a polynomial dependence on the system size, deep in
the ferromagnetic phase, to
slower than polynomial  in the spin-glass phase.
\end{abstract}
{\rm }
\maketitle
\section{Introduction}

Recent years have seen an increasing interaction between the fields of
combinatorial optimization and statistical 
physics\cite{hayes1997,nature1999,TCS2001}.  On one hand,
several problems in the statistical physics of disordered systems have
been mapped onto combinatorial problems, for which
fast combinatorial optimization algorithms are available
\cite{duxbury,opt-phys2001}.  This has provided valuable insights into
questions that are hard to investigate with traditional techniques,
such as Monte Carlo simulations.
On the other hand, concepts and methods from statistical physics are
increasingly applied to combinatorial optimization \cite{TCS2001}.

Easy/hard thresholds analogous to phase transitions have
been observed in random instances of optimization and decision
problems, including satisfiability ({\em SAT}\/)
\cite{MiSeLe,MoZe}, vertex-cover \cite{cover} ({\em VC}\/),
number partitioning \cite{mertens}, and others.
There is currently
much interest in understanding how phase transitions affect the performance 
of
combinatorial algorithms, following the observation\cite{cheeseman}
that the {\em average}\cite{averagecase} or {\em typical} (i.e.
median) running time of some algorithms exhibits a sharp change in
the vicinity of a phase transition.
For example, in 3SAT, a special case of SAT, and in VC, the
typical running time of exact backtracking
algorithms changes\cite{cocco2001,cover-time}
from a polynomial dependence on the input size
in the ''solvable'' region, to exponential dependence in the 
''unsolvable'' 
region.
This provides an insight into the performance of algorithms that goes
beyond the {\em worst-case} running time traditionally considered
in complexity theory. (Note, however, that from the behavior of
individual  algorithms, 
strictly speaking, one cannot draw conclusions about the
``typical hardness'' of a problem itself). 
Recently, statistical physics techniques have been fruitfully
applied to study easy/hard transitions and algorithmic
 performance
\cite{TCS2001}.

In this paper, we apply a branch-and-cut algorithm, a technique
developed in combinatorial optimization, to find the ground state
of the Ising spin glass on random graphs with fixed coordination
number (also called Bethe lattices).

Our motivation is twofold. The first goal is algorithmic: we want to
characterize the typical running time of our
algorithm, notably its behavior across the zero-temperature
spin-glass/ferromagnet phase transition that occurs
when varying the mean of the random couplings. The interest of this
stems  from the importance of branch-and-cut as a
general technique in combinatorial optimization, and from the fact that
finding the ground state of a spin glass is a prominent
example of a hard optimization problem arising from statistical
physics (in general,
it is {\em NP-hard}\/\cite{barahona1982}, see Section \ref{sec:numerics}). The
performance of branch-and-cut for this application has not been
investigated in detail before (see, however,
Refs.~\onlinecite{simone96}, \onlinecite{excited2d}
and \onlinecite{bulk_perturb}), and here we
fill this gap. An interesting aspect is that, unlike in SAT, VC and
other classical combinatorial problems,
here averaging over random instances is physically motivated.
To our knowledge, the only other study relating a ``physical'' phase
transition to algorithmic performance
is that of Middleton\cite{middleton2002}, which
investigates the typical running time of the matching algorithm
for the random-field Ising
model, which however is polynomial everywhere in the parameter space.

We find that the median running time of our algorithm
varies sharply near the
spin-glass/ferromagnet transition, indicating a change from polynomial
time deep in the ferromagnetic phase, to slower than polynomial in the
spin glass phase. We also observed a similar behavior for spin
glasses on regular lattices in two and three dimensions, but will
not report it here.

The second motivation for the present work lies in the ground-state
properties of the Bethe-lattice spin glass, which recently have
attracted a renewed interest
\cite{mezard2001,mezard2002,parisi-tria,boettcher}. Using
branch-and-cut, we compute the ground state energy and magnetization,
and locate the spin-glass/ferromagnet phase transition. This
provides a useful test of recently developed analytical methods to
treat diluted spin glass models
\cite{mezard2001,mezard2002,franz2002}. We solve the
model in the Bethe-Peierls (BP) approximation (equivalent to the
replica symmetric approximation in the replica formalism) using a
stochastic approach proposed in Refs.\onlinecite{mezard2002,thouless2}.
By comparing the branch-and-cut results with the BP results,
 we estimate the magnitude of the replica
symmetry breaking corrections to the ground state energy
and magnetization, finding that they are small.

The rest of the paper is organized as follows. In Section
\ref{sec:model}, we introduce the Bethe-lattice spin glass model.  In
Section \ref{sec:numerics}, we describe the branch-and-cut algorithm
used to calculate the exact ground states of the model. In Section
\ref{sec:bpw}, we describe the Bethe-Peierls approximation and the
stochastic procedure used to solve it. In Section \ref{sec:results}, we
present our branch-and-cut and BP results for the ground state energy and the
zero temperature phase transition.
In Section \ref{sec:complexity}, we show that this transition coincides
with a change of the typical running time. Finally, Section
\ref{sec:discussion} summarizes our results.

\section{Model}
\label{sec:model}

The system considered consists of $N$ Ising spins $S_i=\pm 1$ sitting
on the nodes of a graph $G=(V,E)$, where $V=\{1,\ldots,N\}$ is the set
of nodes and $E=\{(i,j)\}\subset V\times V$ is the set of edges of the
graph.  The energy of the system is given by
\begin{equation}
{\cal H} = - \sum_{(i,j)\in E} J_{ij} S_i S_j
\label{eq:ham}
\end{equation}
where the couplings $J_{ij}$ are  independent, identically distributed
random variables drawn from a Gaussian
distribution $P(\cdot)$ with mean $\mu$ and unit variance,
\begin{equation}
P(J) = \frac{1}{\sqrt{2\pi}} \exp [ -(J-\mu)^2/2 ]
\label{eq:pj}
\end{equation}

We consider the case in which $G$ is a random graph with fixed
connectivity $z$, or $z$-regular graph, where each spin interacts with
exactly $z$ neighbors. This provides a convenient realization of a
Bethe lattice, which avoids some complications associated to the usual
construction of a Bethe lattice from a Cayley tree
\cite{mezard2001}.
Frustration is induced by large loops, the typical size of a loop
being of order $\log(N)$. Small loops are rare, giving the graph a
local tree-like structure, and therefore
the mean field approximation is exact.  A
related model is the Viana-Bray model \cite{viana1985}, in which the
connectivity is a Poisson variable with finite mean.
Finite-connectivity or ``diluted'' models  provide a
better approximation to finite-dimensional spin glasses than
the infinitely-connected Sherrington-Kirkpatrick model.
Furthermore, they are directly related to optimization
problems such as graph partitioning and coloring.

Although it has long been known that replica symmetry is broken in
these two models \cite{viana1985,mottishaw,thouless}, until recently a
replica symmetry broken solution could be found only in some limit
cases. M\'ezard and Parisi recently introduced
\cite{mezard2001,mezard2002} a ``population dynamics'' algorithm which
allows a full numerical solution at the level of one step of replica
symmetry breaking. Explicit results were derived for the Bethe-lattice
spin glass with the  symmetric $\pm J$ disorder distribution, but not for 
the Gaussian distribution considered here or for a non-zero mean.
Previous numerical studies of this model can be found in
Refs.~\onlinecite{lai,boettcher,banavar1987}. For a complete discussion
of the Bethe-lattice spin glass, see Ref.~\onlinecite{mezard2001} and
references therein.

\section{Branch-and-cut algorithm}
\label{sec:numerics}

The problem of
finding a ground state of the Hamiltonian in Eq.~(\ref{eq:ham})
is in general computationally demanding. For a generic graph $G$,
it is NP-{\em hard}\/\cite{barahona1982,opt-phys2001}.
For NP-hard problems, currently
only algorithms are available, for which the running time increases
faster than any polynomial in the system size, in the worst case
(see Section VI for a brief description of
complexity classes). In the special case of a planar system without
magnetic field, e.g.~a square lattice with periodic boundary
conditions in at most one direction, efficient polynomial-time
matching algorithms\cite{bieche1980} exist. For the square lattice
with periodic boundaries in {\em both}\/ directions, 
polynomial algorithms exist
for computing the complete partition function for the $\pm J$
distribution\cite{saul} and for the case in which the coupling
strenghts are bounded by a polynomial in the system
size\cite{GLV2001}. In practice, both algorithms can only reach
relatively small system sizes. 

For the Bethe lattice considered here (and for regular
lattices in dimension higher than two), no polynomial algorithm
is known.
Heuristic algorithms recently used include simulated annealing
\cite{kirkpatrick1983}, ``multicanonical'' simulation\cite{SG-berg92},
genetic algorithms\cite{SG-pal1996b, py}, extremal
optimization\cite{boettcher}, a hierarchical renormalization-group
based approach\cite{houdayer1999}, and the cluster-exact approximation
algorithm\cite{alex-stiff}.
Here, however, we are interested in investigating the running time of
an exact, deterministic algorithm, since in this case the running time
to find the exact ground state is a well defined quantity.
We study the branch-and-cut method\cite{simone95,simone96}
(see Ref.~\onlinecite{JN2001} for a
tutorial on optimization problems and techniques,
including branch-and-bound and
branch-and-cut), which is currently the
fastest exact algorithm for computing spin glass ground 
states\cite{bulk_perturb}, with
the exception of the polynomial-time special cases mentioned above.
As the branch-and-cut method is basically
branch-and-bound with cutting planes, we also did some experiments
with a pure branch-and-bound algorithm\cite{hartwig84,klotz},
which however can only deal with much smaller system sizes,  finding
a qualitatively similar, but less pronounced, variation of the
running time across the transition.

In the rest of this Section, we repeat
a short description of the
branch-and-cut method already given in Ref.~\onlinecite{bulk_perturb},
to the benefit of the reader.
It is convenient to map the problem of minimizing the Hamiltonian in
Eq.(\ref{eq:ham}) into a {\em maximum cut}\/ problem.
Consider a
graph $G=(V,E)$, and let assign weights $\{K_{ij}\}$ to the edges.
Given a partition of the node set $V$ into a subset $W\subset V$ and
its complement $V \setminus W$, the {\em cut}\/ $\delta(W)$ associated
to $W$ is the set of edges with one endpoint in $W$ and the other
endpoint in $V \setminus W$, namely $\delta(W) = \{(ij) \in E \mid i
\in W, j \in V \setminus W \}$. The {\em weight} of $\delta(W)$ is
defined as the sum of the weights of the cut edges, $\sum_{(ij) \in
  \delta(W)} K_{ij}$. The {\em maximum cut} is a node partition with
maximum weight among all partitions. It can be shown \cite{simone95}
that minimizing the Hamiltonian in Eq.(\ref{eq:ham}) is equivalent to
finding a maximum cut of $G$ with the assignment $K_{ij}=-J_{ij}$.

The branch-and-cut algorithm solves the maximum cut problem through
simultaneous lower and upper bound computations. By definition, the
weight of any cut gives a {\em lower bound}\/ on the optimal cut
value. Thus, we can start from any cut and iteratively improve the
lower bound using
deterministic heuristic rules (local search and other specialized
heuristics, see Ref.~\onlinecite{juen-vlsi1988} for details).
How do we decide when a cut is optimal? This can be done by
additionally maintaining {\em upper bounds}\/ on the value of the
maximum cut. Upon iteration of the algorithm, progressively tighter
bounds are found, until optimality is reached.

Since the availability of upper bounds marks the difference between a
heuristic and an exact solution, we now summarize how the upper bound
is computed (for more details, see Ref.~\onlinecite{juen-vlsi1988}.)
To each edge
$(ij)$ we associate a real variable $x_{ij}$ and to each cut
$\delta(W)$ an {\em incidence vector} $\chi^{\delta(W)} \in \R^{E}$
with components $\chi_{ij}^{\delta(W)}$ associated to each edge
$(ij)$, where $\chi_{ij}^{\delta(W)}=1$ if $(ij)\in \delta(W)$ and
$\chi_{ij}^{\delta(W)}=0$ otherwise. Denoting by $P_C(G)$ the convex
hull of the incidence vectors,
it can be shown
that a {\em basic optimum solution}\cite{chvatal1983} of the linear program
\begin{equation}
  \label{eq:mc-formulation}
\max \{ \sum_{(ij) \in E} J_{ij}x_{ij} \mid x \in P_{C}(G)\}.
\end{equation}
is a maximum cut. In order to solve (\ref{eq:mc-formulation}) with
linear programming techniques
we would have to express $P_{C}(G)$ in the form
\begin{equation}
  P_{C}(G) = \{x \in \R^{E} \mid Ax \leq b, 0 \leq x \leq 1\}
\end{equation}
for some matrix $A$ and some vector $b$. Whereas the existence of $A$
and $b$ are theoretically guaranteed, even
subsets of $Ax \leq b$ known in the literature contain a huge number
of inequalities that render a direct solution of
(\ref{eq:mc-formulation}) impractical.

Instead, the branch-and-cut algorithm proceeds by optimizing over a
{\em superset}
$P$ containing $P_{C}(G)$, and by iteratively tightening $P$,
generating in this way progressively better upper bounds. The
supersets $P$ are generated by a {\em cutting plane}\/
approach. Starting with some $P$, we solve the linear program
$\max \{ \sum_{(ij) \in E} J_{ij} x_{ij} \mid x \in P \}$
by Dantzig's simplex algorithm\cite{chvatal1983}. Optimality is proven
if either of two conditions is satisfied: (i) the optimal value equals
the lower bound; (ii) the solution vector $\bar{x}$ is the incidence
vector of a cut.

If neither is satisfied, we have to tighten $P$ by solving
the {\em separation problem}. This
consists in identifying inequalities that are valid for all points in
$P_{C}(G)$, yet are violated by $\bar{x}$, or reporting that no such
inequality exists. The inequalities found in this way are added to the
linear programming formulation, obtaining a new tighter partial system
$P'\subset P$ which does not contain $\bar{x}$. The procedure is then
repeated on $P'$ and so on.

At some point, it may happen that (i) and (ii) are not satisfied, yet
the separation routines do not find any new cutting plane. In this
case, we {\em branch} on some fractional edge variable $x_{ij}$ (i.e.
a variable $x_{ij} \not \in \{0,1\}$), creating two subproblems in
which $x_{ij}$ is set to 0 and 1, respectively. We then we apply the
cutting plane algorithm recursively for both subproblems.

\section{Bethe-Peierls approximation}
\label{sec:bpw}

We recall here the zero-temperature formulation of the BP
approximation, loosely following Ref.~\onlinecite{mezard2002}.
We consider the Hamiltonian Eq.(\ref{eq:ham}) on a random graph with
fixed connectivity $z=k+1$, in which however some spins $S_i$
({\em cavity} spins) have only $k$ neighbors. The
random couplings are drawn from a distribution $P(J)$. The BP
approximation consists in assuming that the ground state energy of
this system is given by $E = const. - \sum_{i} h_i S_i$, where the sum
runs over all cavity spins.
The cavity fields $h_i$, implicitly defined by this relation,
are independent, identically distributed
random variables when considered as a function of the random
couplings. Their distribution $P(h)$ is the central object of
interest, and satisfies a recursion relation derived as follows.
Suppose we add a new spin $S_0$ to the system, which interacts with
$k$ pre-existing cavity spins $S_1,\dots, S_{k}$ through couplings
$J_1,\dots, J_k$, and we minimize the energy with respect to
$S_1,\dots, S_{k}$. Now $S_0$ is a cavity spin, and  it is
easily shown that its cavity field $h_0$ is given by
\begin{equation}
h_0=\sum_{i=1}^{k} u(J_i,h_i) \ ,
\label{h0}
\end{equation}
where $u(J_i,h_i) = {1 \over 2} \left( \vert h_i+J_i \vert - \vert
  h_i-J_i\vert \right)$. This provides a recursion relation for $P(h)$
as the $J$'s fluctuate according to $P(J)$.

Given a spin $S_0$ interacting with $k+1$ neighbors
with couplings $J_1,\dots, J_{k+1}$, the internal field $H$ acting on $S_0$ 
is
\begin{equation}
H=\sum_{i=1}^{k+1} u(J_i,h_i) \ .
\label{hreal}
\end{equation}
Therefore, if we know $P(h)$ we can determine the probability
distribution $P(H)$, and the magnetization
\begin{equation}
m_{BP} = \lim_{\epsilon \to 0^+} \int dH \, P(H) \, \mbox{sgn}(H) \, ,
\label{mbp}
\end{equation}
where $\mbox{sgn}(x)$ is the sign function and $\epsilon$ is
a small field that breaks the symmetry of Eq.(\ref{h0})
with respect to changing the sign of all cavity fields.

The knowledge of $P(h)$ is also sufficient to determine the ground
state energy of the system. As shown in Ref.~\onlinecite{mezard2002},
this can be expressed as
\begin{equation}
e_{BP}=[ \Delta E^{(1)} ] - \frac{k+1}{2} [ \Delta E^{(2)} ]\,,
\label{ebp}
\end{equation}
where $[\cdots ]$ is the expectation value with respect to $P(J)$ and
$P(h)$, and
the quantities $\Delta E^{(1)}, \Delta E^{(2)}$ are given by
\begin{equation}
\Delta E^{(1)} = - \sum_{i=1}^{k+1} a(J_i,h_i) - \vert \sum_{i=1}^{k+1} 
u(J_i,h_i) \vert
\label{e1}
\end{equation}
and
\begin{equation}
\Delta E^{(2)} = - \max_{S_i, S_j} (h_i S_i + h_j S_j + J_{ij} S_i S_j) \, ,
\label{e2}
\end{equation}
where $a(J_i,h_i) = {1 \over 2} \left( \vert h_i+J_i \vert + \vert
  h_i-J_i\vert \right)$ and $S_i,S_j$ in Eq.(\ref{e2}) are two
randomly chosen cavity spins which we connect with the coupling
$J_{ij}$.

The BP recursion, especially at finite temperature,
has been studied extensively (see
Ref.~\onlinecite{mezard2001} and references therein).  In particular, 
M\'ezard
and Parisi \cite{mezard2002} have given an analytic expression of
$P(h)$ for a binary $P(J)$. Klein et al.\cite{klein} solved the finite
temperature BP recursion for Gaussian couplings with $\mu=0$ in the
vicinity of the spin-glass/paramagnet transition.
No analytical solution has been derived for Gaussian couplings at
$T=0$, to our knowledge, although Klein et al. \cite{klein} derived an
analytic solution near the spin-glass/ferromagnet transition
$\mu=\mu_c$ within the {\em mean random-field} approximation.

Here, we employ the stochastic iterative procedure proposed by
M\'ezard and Parisi\cite{mezard2002} for the more general
one-step replica symmetry broken case (see also
Ref.~\onlinecite{thouless} for a previous application of a similar
method). We consider a population of $\cal{N}$ sites, to which we
associate $\cal{N}$ cavity fields, which are initially assigned at
random (with a small positive bias). We then select $k$ sites at
random, extract $k$ couplings from $P(J)$, compute $h_0$ from
Eq.(\ref{h0}), and assign $h_0$ as the new cavity field of a randomly
chosen site.  We iterate this procedure $\cal{M}$ times per site. After a
certain number of iterations, the distribution $P(h)$ will satisfy
Eq.(\ref{h0}).  At each iteration, by merging $k+1$ randomly
chosen sites we compute the internal field $H$ with
Eq.(\ref{hreal}), and $\Delta E^{(1)}$ with Eq.(\ref{e1}), and
by merging two sites we compute $\Delta E^{(2)}$ with
Eq.(\ref{e2}).  After discarding the first ${\cal{M}}/4$ iterations, by
averaging $\mbox{sgn}(H)$, $\Delta E^{(1)}$ and $\Delta E^{(2)}$ over
the remaining iterations we compute the estimates of $m_{BP}$ and $e_{BP}$
from Eqs.(\ref{mbp}) and (\ref{ebp}), and their
statistical error 
from a binning procedure.  We repeated the procedure
for many values of $\mu$, choosing ${\cal{M}}=10^4$ and $\cal{N}$ between
$10^3$ and $10^5$, the larger population being for $\mu$ near the transition
point $\mu_c$. With ${\cal{N}}=10^5$, the iteration requires about one
hour of computer time.

We note that the BP approximation is known to be wrong,
being equivalent to the replica symmetric solution which is unstable.
We have not attempted to use the generalization of the above procedure
to one step of replica symmetry breaking\cite{mezard2001},
since for the Gaussian case considered
it would require a significant computing time, and since 
the BP approximation gives sufficiently accurate
results for our purposes.

\section{Results}
\label{sec:results}

We have studied the Ising spin glass on random graphs with fixed
connectivity $z=4$ and $z=6$. The instance generator first builds a
random regular graph with the algorithm described in Ref.
\onlinecite{SteWor1999}.
We then assign the couplings $J_{ij}$ according to  the
distribution $P(J)$ in Eq.(\ref{eq:pj}).

Using the branch-and-cut approach we were able to study graph sizes up
to $N=400$  for $z=4$ and $\mu\leq 0.9$, and up to $N=200$ for 
$z=6$ and $\mu\leq 0.7$. For larger values of $\mu$, we considered sizes 
up to $N=1280$. 
The ground states for the smallest systems
can be obtained within a second, while the longest computations lasted
at most one day on a typical workstation.
Incidentally, for Ising spin glasses on a regular grid, specialized
heuristics exist that exploit the grid structure, making it possible
to consider larger system sizes than for the model reported here.
More one timing issues, in relation to the spin-glass/ferromagnet 
phase transition, is presented in Section \ref{sec:complexity}.

All the results were averaged over many samples (a sample, or instance,
 is a realization of the random graph with a realization of the
couplings).
The largest number of samples were considered in the vicinity of the
phase transition, where the fluctuations of the magnetization
are larger.
Near the transition, for sizes $N \le 240$ ($z=4$) and $N
\le 160$ ($z=6$) we computed around 5000 samples for each value
of $\mu$; for $N=400$ ($z=4$) and $N=200$ ($z=6$),
around 500 samples for each value of $\mu$.
For sizes larger than these, we computed
up to 280 samples
for each $\mu$. In the following analysis of the ground state energy
and magnetization, we consider only sizes up to $N = 400$ $(z=4)$
and $N = 200$ ($z=6$), since for larger sizes the statistical error is
quite large. In the analysis of running times we will include all
sizes.

\subsection{Ground state energy}
\label{sec:energy}

\begin{figure}
\begin{center}
  \myscalebox{\includegraphics{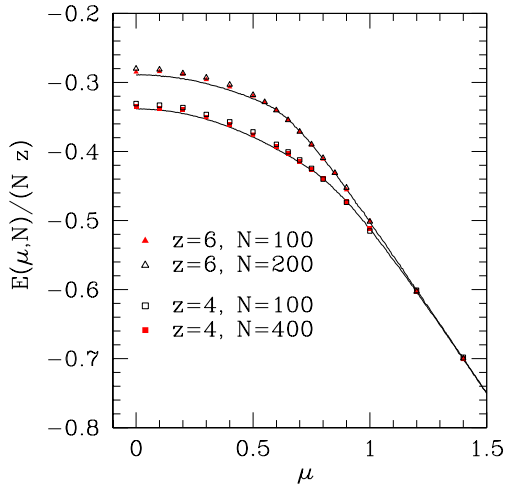}}
\end{center}
\caption{Normalized average ground state energy as a function of 
the mean coupling strength, $\mu$. The symbols
 represent the results of the branch-and-cut
  algorithm. Their statistical errors are smaller than the
  symbol sizes. The lines represent the results of the BP
  recursion. They are obtained by connecting points spaced by
  $\Delta\mu =0.005$ ($\Delta\mu =0.001$ near the transition).  Their
  statistical error is comparable to the line thickness.}
\label{figEnerg}
\end{figure}

We start by showing, in Fig.~\ref{figEnerg}, the average ground state
energy $E(\mu,N)$, divided by $zN$, as a function of $\mu$ for $z=4,6$
and two different system sizes.  For sufficiently large $\mu$, the
system is  completely magnetized, therefore the
ground state energy depends linearly on $\mu$, $E(\mu,N)/N \sim
z \mu$, as visible in the figure.
For small $\mu$ the system is frustrated, hence
the energy saturates. Note that $E(0,N)$
scales as $\sqrt{z}$, not as $z$, therefore the two curves diverge
at small $\mu$.
The lines in Fig.~\ref{figEnerg} represent the numerical solution of
the BP recursion obtained with a population size ${\cal N}=10^3$ 
(we verified that with ${\cal N}=10^5$ the results are unchanged)
and ${\cal M}=10^4$ iterations of the stochastic algorithm.
Clearly, the branch-and-cut results agree well with the BP approximation.

\begin{figure}
\begin{center}
  \myscalebox{\includegraphics{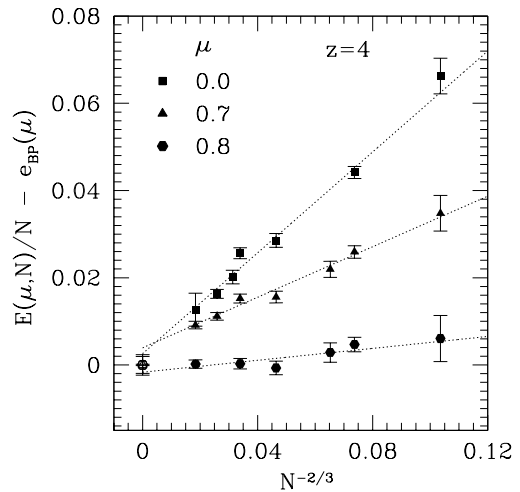}}
\end{center}
\caption{Size dependence of the ground state energy,
  for $z=4$ and different values of $\mu$. The lines represent the best
  fits with the form $E/N = e_\infty + b N^{-2/3}$. The $N=\infty$
  data (origin of the $x-$axis) are obtained in the BP approximation.}
\label{figEnergN_z4}
\end{figure}

We extrapolate the branch-and-cut results to $N=\infty$ by fitting the
data with the form $E/N=e_\infty + b N^{-2/3}$.
As shown in Fig.~\ref{figEnergN_z4},
the finite size corrections are well described by a
$N^{-2/3}$ dependence for small $\mu$, although
an $N^{-\omega}$ correction fits reasonably well the data for
other values of $\omega$ between 0.6 and 1 as well.
For large $\mu$, the finite size corrections are very small.
A  $N^{-2/3}$ correction was also found to fit well
the numerical data by
Boettcher \cite{boettcher}, who computed the average ground state
energy of the $\pm J$ model for $z$ up to $z=26$ and $N$ up to
$N=2048$ using a heuristic algorithm. In Ref.~\onlinecite{mezard2001},
the finite-size dependence of the energy at $T=0.8$ was studied, for
the $\pm J$ distribution and $z=6$, finding a finite-size exponent
$\omega=0.767(8)$, not far from 2/3.  For the Viana-Bray
model with fluctuating connectivity with mean $z=6$,
the value $\omega=0.62 \pm 0.05$, compatible with 2/3,
was found\cite{palassini-phd}, also using a heuristic algorithm.

Fig.~\ref{figEnergN_z4} also shows that the
extrapolated energy, $e_\infty$, is very close to the BP result, $e_{BP}$,
in the whole
range of $\mu$.  Of course, the agreement is not
surprising for large $\mu$, where replica symmetry holds.  For smaller
$\mu$, the observed agreement indicates that replica symmetry breaking
corrections to the ground state energy are small (less than $1\%$).
A similar conclusion was reached in Ref.~\onlinecite{mezard2002}
for the $\pm J$ distribution with zero mean.

In particular, for $\mu=0$ we obtain $e_\infty= -1.38 \pm 0.04$
($z=4$)  and $e_\infty= -1.72 \pm 0.02$
($z=6$), where the errors take into account the uncertainty on the
correction exponent $\omega$, to be compared with our
BP result $e_{BP}=-1.351\pm 0.002$ ($z=4$) and
$e_{BP}= -1.737 \pm 0.002$ ($z=6$).
It is also interesting to compare this with
the ground state energy per spin found in
two\cite{alex-2d-unpublished} and three dimensions\cite{SG-pal1996b}
(which have coordination number $z=4$ and $z=6$, respectively) with
Gaussian couplings and $\mu=0$, which is $e_\infty=-1.31453(3)$ and
$e_\infty=-1.7003(1)$ respectively.

\subsection{Ground state magnetization}
\label{sec:magn}

\begin{figure}
\begin{center}
\myscalebox{\includegraphics{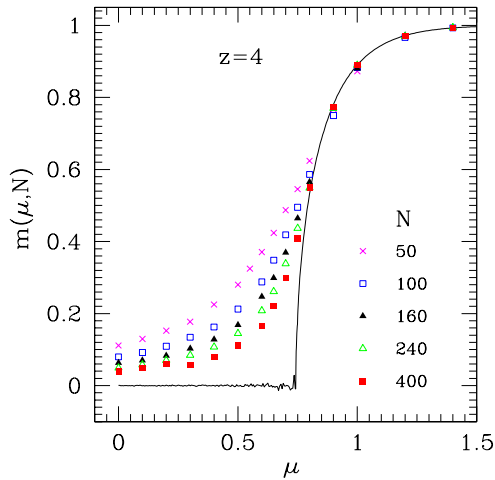}}
\end{center}
\caption{Average ground state magnetization as a function 
of $\mu$, for $z=4$. Symbols: branch-and-cut results
(statistical errors are smaller than the symbols). Line: BP results
 with a population size ranging from ${\cal N}=10^3$ 
(away from the transition) to  ${\cal N}=10^5$ (near the transition),
and with and ${\cal M}=10^4$ iterations of the stochastic algorithm.}
\label{magn_z4}
\end{figure}

\begin{figure}
\begin{center}
\myscalebox{\includegraphics{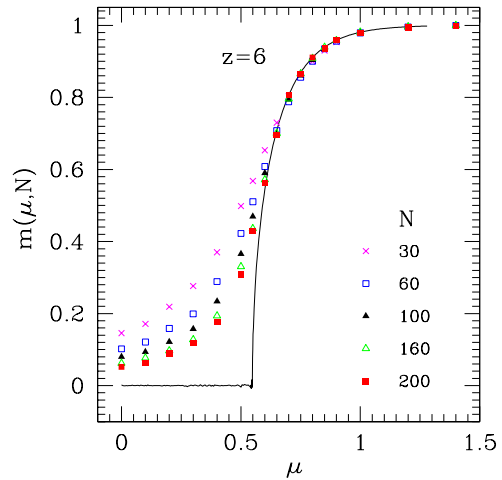}}
\end{center}
\caption{Same as Fig.~\protect\ref{magn_z4} but for $z=6$.}
\label{magn_z6}
\end{figure}

In Figs.~\ref{magn_z4} and \ref{magn_z6} the symbols show, for $z=4$ and $6$
respectively, the average ground state magnetization $m=[M]_J$, where
$M=\frac{1}{N}\sum_i S_i$ and $[\ldots ]_J$ denotes the sample
average, as a function of $\mu$ for different system sizes $N$.  The
lines show the $N=\infty$ result in the BP approximation.  For small
$\mu$, the magnetization vanishes as $1/\sqrt{N}$. For large $\mu$,
the finite-$N$ data agree with the BP result within the error bars,
with negligible finite-size corrections (again, we recall that the BP
approximation
is exact for sufficiently large
$\mu$, hence the agreement is expected). 
From the point at which the BP magnetization vanishes, we
estimate the critical coupling strength
$\mu_c=0.742 \pm 0.005$ ($z=4$) and $\mu_c=0.546 \pm 0.005$
($z=6$).  Note that recursion relation Eq.(\ref{h0})
admits two symmetric solutions for $\mu > \mu_c$.  Hence, in the
stochastic procedure the magnetization will oscillate between positive
and negative values, with an oscillation ``time'' (number of
iterations) ${\cal{M}}_0$ that increases with the population size $\cal{N}$
and with $\mu$.  Therefore, to compute the magnetization
correctly, we need ${\cal{M}}_0 \gg {\cal{M}}$.
To do this, we increased the size of the population
progressively from ${\cal N}=10^3$ to ${\cal N}=10^5$ as $\mu$
approached $\mu_c$. (Residual oscillations very close to $\mu_c$
introduce a small systematic error\cite{footsmaller}, which
is reflected in the errors for $\mu_c$ quoted above.)

\begin{figure}
\begin{center}
  \myscalebox{\includegraphics{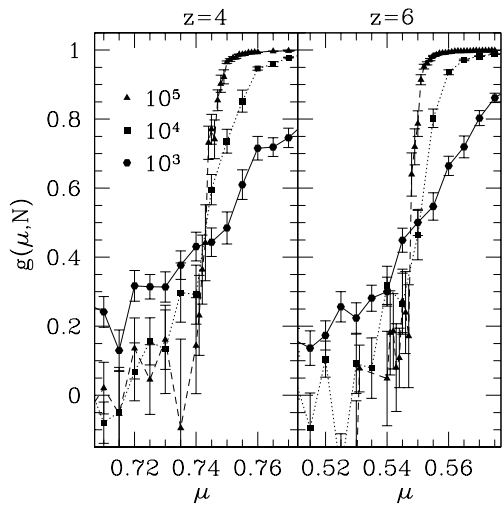}}
\end{center}
\caption{Binder cumulant from the stochastic solution
  of the BP ansatz, for three different sizes of the
  stochastic population ${\cal{N}}$.}
\label{binder_bethe}
\end{figure}

Another estimate of $\mu_c$ 
can be obtained from the Binder cumulant\cite{binder}
\begin{equation}
g(\mu) = \frac{1}{2}
\left( 3-\frac{[ M^4 ] }{[ M^2 ]^2}\right) \, ,
\label{defbinder}
\end{equation}
where $[\cdots ]$ is now the ``time'' average.  In the limit $\cal{N}\to
\infty$, $g(\mu)=0$ for $\mu < \mu_c$ and $g(\mu)=1$ for $\mu >
\mu_c$, hence $g(\mu)$ can be used to locate $\mu_c$. As shown
in Fig.~\ref{binder_bethe}, 
the variation of the Binder cumulant with $\mu$ sharpens as ${\cal N}$ increases, 
an effect of the sign oscillations of the magnetization, which become less
important as ${\cal{N}}$ increases.
From ${\cal{N}}=10^5$ we estimate
\begin{eqnarray}
\mu_c^{BP} &=& 0.743 \pm 0.005 \quad (z=4) \nonumber \\
\mu_c^{BP} &=& 0.547 \pm 0.005 \quad (z=6) \nonumber
\label{bpw_estimate}
\end{eqnarray}
which agrees with the above
estimate from the average magnetization. We also
verified that with these values of $\mu_c$, the magnetization
obeys $m_{BP} = a (\mu-\mu_c)^\beta$ for $\mu
\simeq \mu_c$, with the mean-field exponent $\beta=1/2$ and $a\simeq
0.23$.

Klein et al. \cite{klein} solved the BP recursion in the vicinity of
$\mu_c$ using the  mean random field approximation (MRF).
Their results $\mu_c^{MRF} = 0.775$ ($z=4$) and $\mu_c^{MRF} = 0.587$
($z=6$) (obtained after rescaling their value by an appropriate
normalization factor $\sqrt{z}$) are slightly larger than
our result $\mu_c^{BP}$.

\begin{figure}
\begin{center}
\myscalebox{\includegraphics{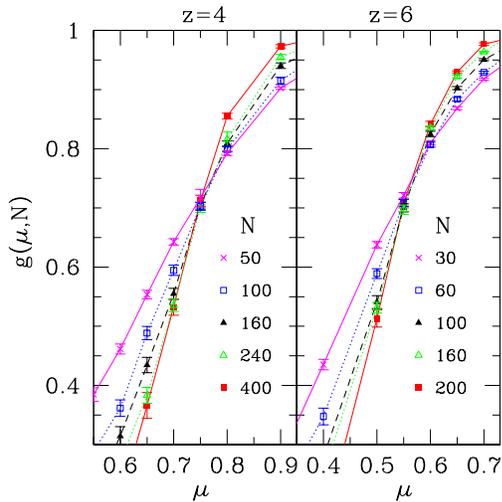}}
\end{center}
\caption{Binder cumulant as a function of $\mu$ for various
system sizes. Only the region around the
  phase transition is shown. The lines are only a guide to the eye.}
\label{figBinderZoom}
\end{figure}

In order to obtain an estimate of $\mu_c$ from the finite-$N$
branch-and-cut data,
we computed the Binder cumulant $g(\mu,N)$, defined as in
Eq.(\ref{defbinder}) but with the time average replaced by the sample
average.  According to finite-size scaling, the curves for $g(\mu,N)$
as a function of $\mu$ for various $N$ must cross at the critical
point $\mu=\mu_c$.  In Fig.~\ref{figBinderZoom} we plot the Binder
cumulant in the vicinity of the intersection point (note that the
horizontal scale is much larger than that of Fig.~\ref{binder_bethe}),
from which we obtain
\begin{eqnarray}
\mu_c &=& 0.77 \pm 0.02 \quad (z=4) \nonumber \\
\mu_c &=& 0.56 \pm 0.02 \quad (z=6) \nonumber .
\end{eqnarray}

\begin{figure}
\begin{center}
  \myscalebox{\includegraphics{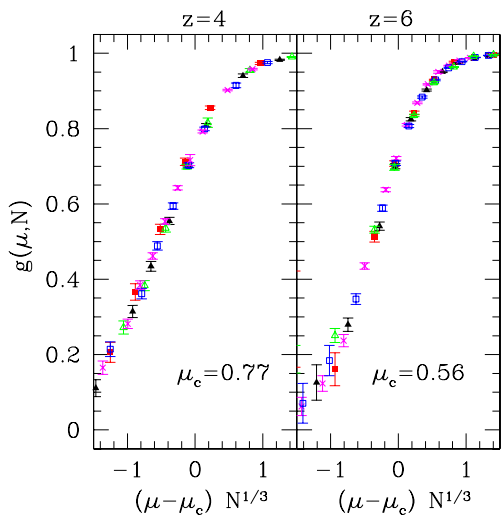}}
\end{center}
\caption{Scaling plot for the Binder cumulant. The symbols 
 are the same as in the corresponding panels in Figure
  \ref{figBinderZoom}. 
 Note the steeper shape of the scaling function for $z=6$.}
\label{figScalingBinder}
\end{figure}

This agrees with $\mu_c^{BP}$ within the error bars, suggesting
that also for the magnetization replica symmetry breaking corrections
are small, causing a shift of $\mu_c$ of less than $3 -
4\%$. Replica symmetry breaking corrections are expected to increase
with $z$.  In the
Sherrington-Kirkpatrick model (which is the $z\to\infty$ limit of the
present model), corrections shift $\mu_c$ from 1.25 to 1. Although our
numerical estimate of $\mu_c$ is slightly {\em larger} than
$\mu_c^{BP}$ instead,
this could be a statistical fluctuation or a finite-size
effect. The small size of replica-symmetry-breaking corrections
to $\mu_c$ suggests that the mixed ferromagnetic spin-glass
phase is narrow for these values of $z$,
as also recently indicated for the three-dimensional
Ising spin glass\cite{martin_mixed}.

The Binder cumulant is expected to satisfy the following
finite-size scaling relation\cite{privman} for $\mu \simeq \mu_c$:
\begin{equation}
g(\mu,N)=\tilde{g}(N^{1/(d_u\nu)}(\mu-\mu_c))\,
\end{equation}
where $d_u$ is the upper critical dimension, which for the Ising spin
glass is $d_u=6$.
As usual, by plotting $g(\mu,N)$ against $N^{1/(d_u\nu)}(\mu-\mu_c)$
with correct parameters $\mu_c$ and $\nu$, the data points for
different system sizes should collapse onto a single curve near
$(\mu-\mu_c)=0$.   As shown in
Fig.~\ref{figScalingBinder}, using the estimates of $\mu_c$ obtained
above and the mean-field exponent $\nu=1/2$ we obtain a good data
collapse, showing that finite size scaling is well satisfied in our
range of sizes.

\begin{figure}
\begin{center}
  \myscalebox{\includegraphics{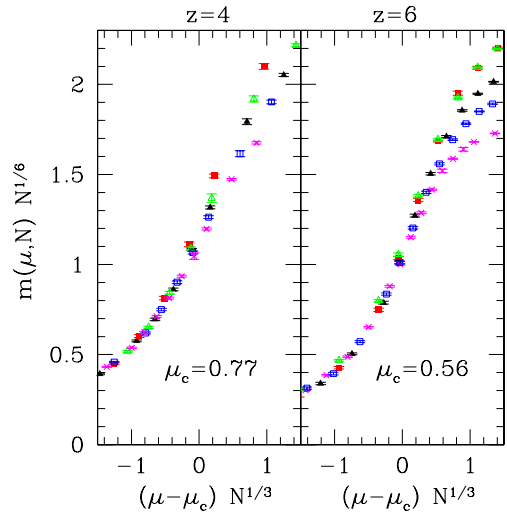}}
\end{center}
\caption{Scaling plot for the ground state magnetization. The symbols 
   are the same as in Fig.~\ref{figBinderZoom}. }
\label{figScalingMag}
\end{figure}

In Fig.~\ref{figScalingMag} we also show scaling plots for
the average magnetization $m(\mu,N)=[M]_J$, whose scaling form is
\begin{equation}
m(\mu,N)=N^{-\beta/(d_u\nu)}\tilde{m}(N^{1/(d_u\nu)}(\mu-\mu_c))\,,
\label{eq:scalingMag}
\end{equation}
with the mean field exponent $\beta=1/2$. The data show a good scaling
collapse for $\mu\le \mu_c$.

\section{Typical running time of our branch and
  cut algorithm}
\label{sec:complexity}

In this section we study the running time of our program as a function
of the mean coupling strength $\mu$. In computer science the
complexity of a problem is classified in terms of the {\em worst-case}
running time of its solution algorithms\cite{GaJo,mertens_lecture}.  
Central notions here
are the complexity classes P and NP. Informally,
the class P consists of all {\em decision}\/
problems (namely, problems whose solution can only be ``yes'' or
``no'') for which at least one algorithm is known that can {\em
  generate}\/ an answer in polynomial time, even in the ``worst case''.
The class NP consists of all decision problems for which,
if for a given instance the answer is ``yes'', then 
there is a {\em certificate}\/ from which the
correctness of the answer can be {\em verified}\/ in polynomial time. For
example, the question ``Given a spin glass instance, is there a spin
configuration with energy less than or equal to $E_{0}$?'' belongs to
NP. If for a given instance the answer is positive, then there is 
a spin configuration with correct energy, and its correctness 
can be verified in polynomial time. 
Only the {\em existence}\/ of such a certificate (spin
configuration, in the above example) is required,  not the ability
 to {\em find}\/ it in polynomial time. 
The class NP contains P, but it might be larger 
(many believe it is larger, and answering the question whether P = NP is an
important open problem). 
NP-{\em complete} problems are the ``most difficult'' in the class NP, 
in the sense that no polynomial algorithm is known for solving them, 
and if a 
polynomial algorithm could be found for {\em one} of them, 
this would imply that {\em all} of them are polynomially
solvable\cite{ACGKMP1999}. 
The classes P and NP are defined for decision problems, but similar 
ideas apply to combinatorial optimization problems as well.
Informally, an optimization problem is called NP-{\em hard}, if it is
at least as difficult as every NP-complete problem. In particular, an
optimization problem 
is NP-hard if the associated decision problem
is NP-complete. This is true for many optimization problems, e.g. the
maximum cut problem or the travelling salesman problem.

In practice, the running time
can vary greatly from an instance of the problem to another,
and the worst-case running time might very rarely occur. Recent work
has therefore focused on the {\em average}  running
time with respect to  random instances
drawn from some probability distribution. Instead of the average one
can also analyze the median, or {\em
  typical} running time,  which
has the advantage of being less
influenced by the occurrence of exponentially rare samples with
huge running times.

It should be noted that, unlike the worst-case
complexity classification discussed above, which is an
algorithm-independent feature of the problem itself, in general the typical
running time can be different for different algorithms and
implementations that solve the same problem.

Returning to our problem, as mentioned in Section III, finding the
ground state of the Bethe-lattice spin glass is an NP-hard problem.
For {\em all} values of $\mu$ the {\em possible}
realizations of the disorder are the same as for $\mu=0$.  Hence,
the algorithm has an exponential worst-case
running time  even on instances deep in the ferromagnetic phase.
However, for large $\mu$ highly frustrated
realizations are very unlikely to appear, hence the {\em typical}
running time will decrease as $\mu$ increases.  The question we ask
here is whether, for large $N$, the running time undergoes a {\em sharp}\/
transition as a function of $\mu$ and, if so, whether the transition
coincides with the spin-glass/ferromagnet phase transition.

One may use the CPU time as a measure of the running time. However,
the CPU time is machine-dependent, hence it is not suitable when
different computers are used. Furthermore, it is hard to separate out the
influence of size-dependent hardware effects on the CPU time  (for example,
 small problems can be fully stored in the cache and
therefore run faster).
To avoid these problems, in the following we use the {\em number of
linear problems solved}, $n_{lps}$, as a measure of the
running time\cite{bulk_perturb}.

\begin{figure}
\begin{center}
  \myscalebox{\includegraphics{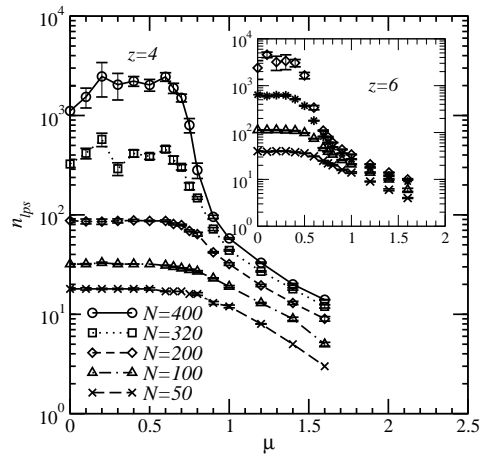}}
\end{center}
\caption{Median running time (measured in number of solved linear problems)
  as a function of $\mu$ for different system sizes. 
Inset shows the same for $z=6$ and $N=200,140, 100,50$.}
\label{figLpsMu}
\end{figure}

In Fig.~\ref{figLpsMu} we show the median 
running time  so defined as a
function of $\mu$ for $z=4,6$ and different system sizes.
Clearly, 
 ground states are calculated
quickly in the ferromagnetic region, while in the spin-glass phase the
 running time increases dramatically (note the logarithmic scale on the vertical axis), 
and is approximately constant
within the entire spin glass phase. The variation becomes more
pronounced as $N$ increases, suggesting a sharp discontinuity
in the $N\to\infty$ limit
around $\mu \approx 0.8$ ($z=4$) and $\mu \approx 0.6$ ($z=6$), which
is close to the spin-glass/ferromagnet transition point $\mu_c$ 
determined in Section \ref{sec:results}.

\begin{figure}
\begin{center}
  \myscalebox{\includegraphics{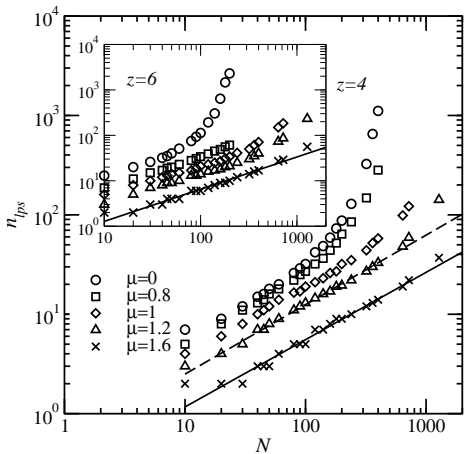}}
\end{center}
\caption{Median running time 
  as a function of $N$ for different coupling strengths
  $\mu$ in logarithmic scale. The straight
  lines represent power-laws $c*N^{\zeta}$ with $\zeta=0.699$ ($z=4,
  \mu=1.2$), $\zeta=0.677$ ($z=4,\mu=1.6$) and $\zeta=0.709$
  ($z=6,\mu=1.6$), respectively, showing that in the ferromagnetic
  phase the median running time is polynomial.}
\label{figLpsN}
\end{figure}

As shown in Fig.~\ref{figLpsN}, deep in
the ferromagnetic phase the data is consistent with a polynomial
increase of the running times with $N$.  For smaller values of $\mu$, the
curves are bending upwards, indicating that the running time increases
faster than any polynomial. This is also the case for $\mu=0.8$ ($z=4$) and
for $\mu=0.6$ ($z=6$, not shown). Hence from this data, it seems that
the change in the typical running time  
 occurs at a value of $\mu$ larger than
$\mu_c$, although it is difficult
to locate a precise transition point.
A mismatch between phase transition and change of
the running time
has been observed before, e.g. for a simple algorithm solving vertex
cover\cite{cover-time}.

We have fitted the data in Fig.~\ref{figLpsN}
with a function of the form $n_{lps}(N)\sim
\exp({bN^c})$. For $\mu=0$, we find $b=0.026(9)$, $c=0.87(5)$ for
$z=4$, and $b=0.007(3)$ and $c=1.24(8)$ for $z=6$, but the data
exhibits in both cases a considerable scatter around the fitting
region, prohibiting to conclude in a definite way that the typical
running time is exponential.  Nevertheless,  the data strongly 
suggest so.

\section{Conclusions}
\label{sec:discussion}

We have studied the ground state of a diluted mean-field Ising spin
glass model with fixed connectivities $z=4, 6$ and Gaussian distribution of
the couplings, with mean $\mu$ and unit variance. We have applied a
branch-and-cut algorithm, a sophisticated technique originating in
combinatorial optimization which guarantees to find exact ground
states. Our motivation was to study the  spin-glass/ferromagnet
transition and relate it to the change in the typical running time of
our algorithm.

From the study of the Binder cumulant, we have obtained values for the
critical  coupling strength, $\mu_c$.
We have also solved the model in
the Bethe-Peierls approximation, using an iterative stochastic
procedure. In this approximation we obtain a critical  coupling strength,
$\mu_c^{BP}$,
which agrees with the branch-and-cut estimate within the error bars
of the latter, indicating that replica symmetry breaking effects are
quantitatively small. Finite-size scaling is well satisfied for systems
of size larger than $N \approx 30$.

We have also analyzed the ground state energy, and shown that the
branch-and-cut results, extrapolated to the thermodynamic limit, are
in very good agreement with the Bethe-Peierls results, again indicating that
replica symmetry breaking effects are quantitatively small. In the spin 
glass region,
finite-size corrections are well described by a $N^{-2/3}$ dependence.

We have investigated the typical running time of  our implementation of the 
branch-and-cut
algorithm, which we defined as the median number of linear programs
needed to find the ground state,  with respect to a uniform distribution
over the space of instances.
We have shown that finding ground states is ``hard'' 
in the spin-glass phase, and ``easy'' 
deep in the ferromagnetic region, with a sharp variation
at a value of $\mu$ slightly larger than $\mu_c$. 
The data indicate that while the worst-case running time is always
exponential in the system size, the typical running-time is 
polynomial in the ferromagnetic phase and super-polynomial in the spin-glass
phase.

Our understanding of what makes a problem computationally hard is
still very weak. In this paper, we have shown that in a standard
hard problem from physics, the Ising spin glass, 
a ``physical'' phase transition has a dramatic effect on
the performance of a solution algorithm.
Although in principle the ``typical hardness'' 
is algorithm-dependent, it is reasonable to expect that 
the phase transition will influence to some extent 
the running time of many other solution algorithms.
Furthermore, we expect that similar phenomena
occur in other well-known physical models.

\begin{acknowledgments}
We thank A.P. Young for comments on an earlier version of the paper,
and the Regional Centre of Computing in Cologne for the
  allocation of computer time and various support. 
MP also thanks Alan Bray for a useful correspondence. AKH obtained
  financial support from the DFG (Deutsche Forschungsgemeinschaft)
  under grants Ha 3169/1-1 and Zi 209/6-1.
\end{acknowledgments}


\begin{thebibliography}{99}
\bibitem{footsmaller} For these values of $\mu$,
we restricted the average to the time intervals in which
the magnetization is positive.
\end{thebibliography}
\end{document}